\documentclass{aa}
\date{}
\usepackage{graphicx}
\usepackage{natbib}
\usepackage{txfonts}
\def\new#1 {{\bf #1 }}
\def\S{Sect. }

\def\vt{\object{V773\,Tau}\ }

\def\cut#1 {\sout{#1} }
\def\beq{\begin{equation}}
\def\eneq{\end{equation}}
\def\simgt{\lower.5ex\hbox{$\; \buildrel > \over \sim \;$}}
\def\simlt{\lower.5ex\hbox{$\; \buildrel < \over \sim \;$}}
\begin{document}
\title{
 Interacting coronae of two T Tauri stars:
first observational evidence for solar-like helmet streamers}
\author{
M.  Massi\inst{1},
E.  Ros\inst{1},
K. M.   Menten\inst{1},
M. Kaufman Bernad\'o\inst{1}\thanks{Humboldt Research Fellow},
G.  Torricelli-Ciamponi\inst{2},
J.  Neidh{\"o}fer\inst{1},
A.  Boden\inst{3},
D.  Boboltz\inst{4},
A.  Sargent\inst{5}
\and
G. Torres\inst{6}}
\institute{%
Max-Planck-Institut f\"ur Radioastronomie, Auf dem H\"ugel 69, D-53121 Bonn, Germany \\
\email{mmassi, ros, kmenten, mkaufman, jneidhoefer@mpifr-bonn.mpg.de}
 \and
INAF, Osservatorio Astrofisico di Arcetri, Largo E. Fermi 5, I-50125 Firenze, Italy \\
\email{torricel@arcetri.astro.it}
\and
Michelson Science Center, California
Institute of Technology, 770 South Wilson Ave., Pasadena CA 91125, USA\\
\email{bode@ipac.caltech.edu} 
\and
U.S.~Naval Observatory, 3450 Massachusetts Ave, Washington DC 20392, USA\\
\email{dboboltz@usno.navy.mil} 
\and
Division of Physics, Math, and Astronomy, California
Institute of Technology, MS 105-24, Pasadena, CA 91125, USA\\
\email{afs@astro.caltech.edu}
\and
Harvard-Smithsonian Center for Astrophysics, 60
Garden St., Cambridge, MA 02138, USA\\
\email{gtorres@cfa.harvard.edu}
}

\offprints{M. Massi} %, \\ \email{mmassi@mpifr-bonn.mpg.de}}

\abstract
{The young binary system V773 Tau A exhibits a persistent radio flaring activity that gradually increases
from a level of a few mJy  at apoastron to more than  100 mJy  at periastron. 
Interbinary collisions between  very large ($> 15 R_*$) magnetic structures anchored on the  two rotating  stars of the  system
have been proposed
to be the origin of these periodic  radio flares.
Magnetic structures extended over tens of  stellar radii, 
that can also account for the observed fast decay of the radio flares, 
seem to correspond to the typical solar semi-open quite extended 
magnetic configurations  called  helmet streamers.}
{We  aim to find direct observational evidence for the postulated, solar-like, coronal topologies.}
{We performed  seven-consecutive-day  VLBI observations at  8.4 GHz using an array consisting of the VLBA and the 100-m Effelsberg telescope.  \vt A was phase-referenced to QSO B0400+258.}
{Two distintive structures appear in the radio images here presented. 
They happen to be associated with the  
primary and secondary stars  of the \vt A system.
In one  image (Fig.~\ref{vlbi}-B) the two features are  extended up to
$18 R_*$ each and  
are    nearly parallel revealing the presence of two interacting
 helmet streamers.
One  image (Fig.~\ref{vlbi}-E) taken a few hours after a flare
monitored by the 100-m Effelsberg telescope 
shows one elongated fading structure  
substantially rotated with respect to  those  seen in the B run.
The same decay scenario is seen in Fig.~\ref{vlbi}-G
for the helmet streamer associated with the other star.}
{This is the very first direct evidence revealing that even
if the flare origin is magnetic reconnection due to interbinary collision,
 both stars 
independently emit in the radio range with structures of their own. These 
 structures are  helmet streamers,
observed  for the first time in stars other than the Sun.
The complete extent of each helmet streamer
      above the stellar surface is about 
       24$R_*$
      which implies that they can  practically interact throughout the whole orbit, even rather close to
 apoastron where the stellar separation is $52 R_*$.
However, the radio flares become stronger when the stars approach.
Around  periastron
the stellar separation is only  $30 R_*$, nearly covered  by
a single streamer: the two
streamers  
overlap producing  the observed giant  flares.}

\titlerunning{Observational evidence for solar-like helmet streamers}

\authorrunning{M. Massi et al.}

\maketitle

\section{Introduction} \label{introduction}
\begin{figure}[]
\centering
\includegraphics[angle=0, scale=1]{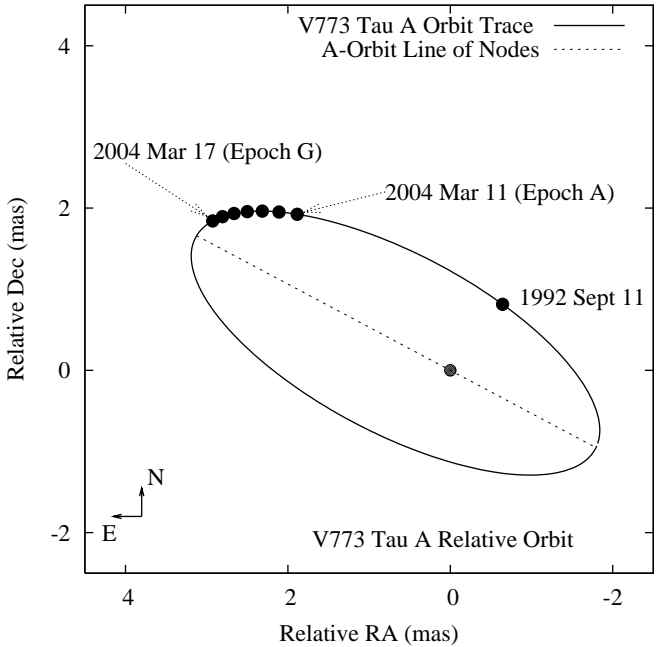}
\caption{
 Orbit of  the young stellar binary system \vt
 based on  the interferometric-spectroscopic  model by Boden and collaborators
 (2007).
The secondary is given at
our seven (A-G) VLBI observations (epoch 2004) and at Phillips and
collaborators  (1996)  VLBI observation (epoch 1992).
Stellar diameters are rendered to scale.
}
\label{boden}

\end{figure}
Located in the  Taurus-Auriga star forming region, at a distance of 136 pc,
the young binary system V773 Tau A (see Fig. ~\ref{boden}) is
formed by two stars of $1.5 M_{\odot}$ and $1.3 M_{\odot}$ (Boden et al. 2007).
This system  is a highly variable radio source: it  exhibits persistent flaring activity that gradually increases from its weakest level at apoastron, few mJy,
to its strongest at periastron, more than 100 mJy.
The strong correlation observed between the flare intensity
and the  distance between the two stars  suggests the occurrence of  interbinary collisions of the binary components' very extended
coronae (Massi et al. 2002, 2006).
Knowing that the distance between the two stars at apoastron and at periastron is $52 R_*$ and $30 R_*$, respectively
($R_*$ stellar radius, $R_*\simeq 2 R_{\odot}$ for both stars,
which corresponds to 0.069 mas at 136 pc, Boden et al. 2007), interbinary collisions would then imply very large coronal structures  with
a size  of $15 R_* \le H \leq 26 R_*$.
On the other hand,  X-ray observations have
provided clear evidence for much smaller coronal loops
with sizes of 0.6--1.4$~R_*$
(Skinner et al. 1997; Tsuboi et al. 1998).
Although the radio and X-ray observations might appear contradictory,
a multiwavelength campaign carried out on  this system (Feigelson et al. 1994)
proved that
the radio and
X-ray emissions come from spatially separated regions.
In fact  this study  showed  that
during radio variability
the X-ray flux remained steady.

This implies that the coronae in the \vt A stellar system  are
formed by at least two different types of magnetic structures: one, stellar
 sized,
 associated with
the X-ray emission and  another, extended, associated with the radio emission.
In the very well studied solar corona it is known that there exist
small, closed coronal loops as well as much more extended semi-open
structures called helmet streamers.
Solar helmet streamers
are formed by a helmet, i.e. a cusp-shaped field line located on top (2 to 4 $R_{\odot}$) of
a  coronal loop, and a very long streamer which extends
out to $30 R_{\odot}$ as can be seen in optical images from the  Solar and Heliospheric Observatory (SOHO) satellite
(Schwenn 2006; Suess \& Nerney 2004; Vourlidas 2006).
Following
the analogy with the Sun,
Massi and collaborators (2006) proposed
a model where  helmet streamers 
are the basic magnetic  topology 
 at the origin of the observed  flares of the \vt A  system. 
This
model
with a  helmet located at 2--$5 R_*$
and an upper  mirror point at the  top of a streamer
of  10--$20 R_*$, was  able to
explain the fading phase of the flares in \vt A in the centimeter  and
millimeter radio band.
In this scenario
the  extended  streamers collide 
 and magnetic reconnection  
takes place far out from the stellar surfaces.
Therefore, the radio flares
in \vt A  depend on magnetic interactions like in the case of the solar flares,
with the difference that the interacting
magnetic structures belong to the two
distinct coronae of each of the stars of the binary system. These interacting magnetic structures
are  streamers, most likely
located at the top of the observed stellar sized X-ray
emitting coronal loops.

 At this point, it is important to mention the large radio emitting structure imaged
 during a Very-Long-Baseline Interferometry (VLBI) observation carried
 out close to the periastron passage of  V773 Tau A  in 1992 by Phillips and collaborators
(1996, see  Fig.~\ref{boden}).
 This VLBI image shows a large  halo ($\ge$ 3 mas)
 surrounding  two compact components that are
 separated by 1.14 mas.
      The two components  have been interpreted by Phillips
and collaborators  as
      emitting zones associated with each of the two stars of the binary system. In fact, the
      separation of 1.14 mas, i.e. $16 R_*$, inferred from their observation is completely
      consistent with the projected stellar separation of 1.03 mas, 
i.e. $15 R_*$ (see Fig. 1).
How is this halo related to the postulated helmet streamers?
Are the streamers  existing as stable independent  magnetic structures
or are they  continuously merged together forming the observed halo?

In order to investigate this complex coronal topology we decided
to perform a long (seven-consecutive-day) set of new  VLBI observations
at a time when the system is quite  displaced from periastron,
so that the larger distances between the two stars
could reduce the  overlap between the streamers and
therefore the  formation of  joint structures.
We describe these new VLBI observations and our results 
in \S \ref{obs} and \S  3. 
In Sect. 3.1 we discuss the position of the two stars of the binary system
 with respect to the radio peaks present in each observation. 
In \S  3.2 we interpret
the resolved structures as helmet streamers, in
\S  3.3 we discuss the possible distribution of helmet streamers
around the stellar surface,
and in Sect. 3.4 we focus on the short-living  upper mirror
points closing the streamers.
In \S  4.1
the different flaring activity at apoastron and periastron,
and  the different observed topologies
are discussed on the basis of the here presented  results.
Finally, in \S 4.2, we discuss the importance of the helmet
streamer configuration in Astrophysics.

\section{Observations and Data Reduction}\label{obs}

We observed \vt A at 8.4\,GHz with an array consisting of the Very Long 
Baseline Array (VLBA, 10 telescopes of 25\,m each across the USA) together 
with the  Effelsberg 100-m radio telescope (VLBA+EB).  The addition 
of Effelsberg provided long baselines and a higher sensitivity, together 
with the possibility of measuring the flux density of the target source 
in between the VLBI scans during the observation. The observing runs, listed 
as segments A to G (code \texttt{BM198}) in the first column of Table 1, 
covered the UT range from 17:30 to 23:00,  on the JD time given in 
the second column. The 3rd column of the table gives the orbital phase 
calculated for the orbital period, $P=51.1033$\,d, and the time periastron 
passage, T$_{0}$=53059.50 MJD (Boden et al. 2007). The selected range of 
orbital phases, $\Phi=0.32$--$0.43$ implies larger distances between the 
two stars close to apoastron (see Fig. ~\ref{boden}). 

The data were recorded at each station 
over four 8-MHz bands at a 
bit rate of 128\,Mbps, in dual polarization with 1-bit sampling, yielding 
a  total bandwidth of 32\,MHz for each polarization hand.  The data were 
processed at the VLBA correlator at the NRAO's Array Operations Center in 
Socorro, New Mexico, using a pre-averaging time of 4\,sec.  We chose as reference 
source the QSO \object{B0400+258} (CTD\,26, J0403+2600), 3.3\degr\ apart 
from the target source \vt A, and switched between both with cycles of 
$\sim$7\,min, observing 360\,sec at the target source and 90\,sec at the 
reference source. 
Before every fourth scan at the target source, 
Effelsberg measured the total flux density at the observing frequency. 

We 
used a catalog position for B0400+258 of 
$\alpha_{J2000.0}=04^\mathrm{h}03^\mathrm{m}05\rlap{.}^\mathrm{s}586056$ and 
$\delta_{J2000.0}=26^{\circ}00^\prime01\rlap{.}^{\prime\prime}50288$. The 
position used for \vt  was 
$\alpha_{J2000.0}=04^\mathrm{h}14^\mathrm{m}12\rlap{.}^\mathrm{s}919833$ and 
$\delta_{J2000.0}=28^{\circ}12^\prime12\rlap{.}^{\prime\prime}19953$, 
after computing the proper 
motion from the reference epoch of 1993.88, with rates of $ \Delta \alpha$ 
0.42 mas/yr and $\Delta \delta -23.25$ mas/yr. We used the QSO 
\object{3C\,84} as fringe finder, and additionally, to check the stability 
of the astrometry we also had blocks of observations of the reference 
source \object{B0405+305} (\object{J0408+3032}).  Polarization calibrators 
were \object{DA\,193} and \object{B0059+581} (\object{J0102+5824}), and 
\object{4C\,28.07} (\object{CTD\,20}, \object{B0234+285}, 
\object{J0237+2848}).

The post-correlation data reduction was performed using the Astronomical 
Image Processing System ($\cal A\cal I\cal P\cal S$). A-priori visibility 
amplitude calibration was done using the antenna gains and the system 
temperatures measured at each station.   Instrumental phase offsets 
between the 8-MHz bands, caused by differing paths through the electronics, 
were determined using short section of data on the source  \object{DA\,193}
and applied to all of the data.  To correct for any residual phase rates and
delays, a fringe-fit was performed on the reference source, QSO \object{B0400+258},
and the resulting solutions were interpolated to the target source data.
Finally, an iterative self-calibration and imaging procedure was 
performed on the reference source to correct for any residual phase 
and amplitude errors.  The resulting corrections were then applied to the 
\vt A data.  The same calibration procedure was repeated for all 
the seven epochs of observations.  With the calibration completed, 512$\times$512
pixel total intensity (Stokes {\em I}) images of the target source were produced 
with a pixel spacing of 0.2~mas.  The synthesized beam sizes for the 7 epochs 
were roughly the same at $\sim$1.4$\times$0.5 mas.  Figure\ \ref{vlbi} shows 
the images of \vt A produced for all the seven  epochs.

\section{Results and Discussion}\label{results}

\subsection{Radio Peaks and Position of the Two Stars}\label{results}
\begin{figure*}[t]
\centering
\resizebox{\hsize}{!}{\includegraphics[angle=0, scale=0.75]{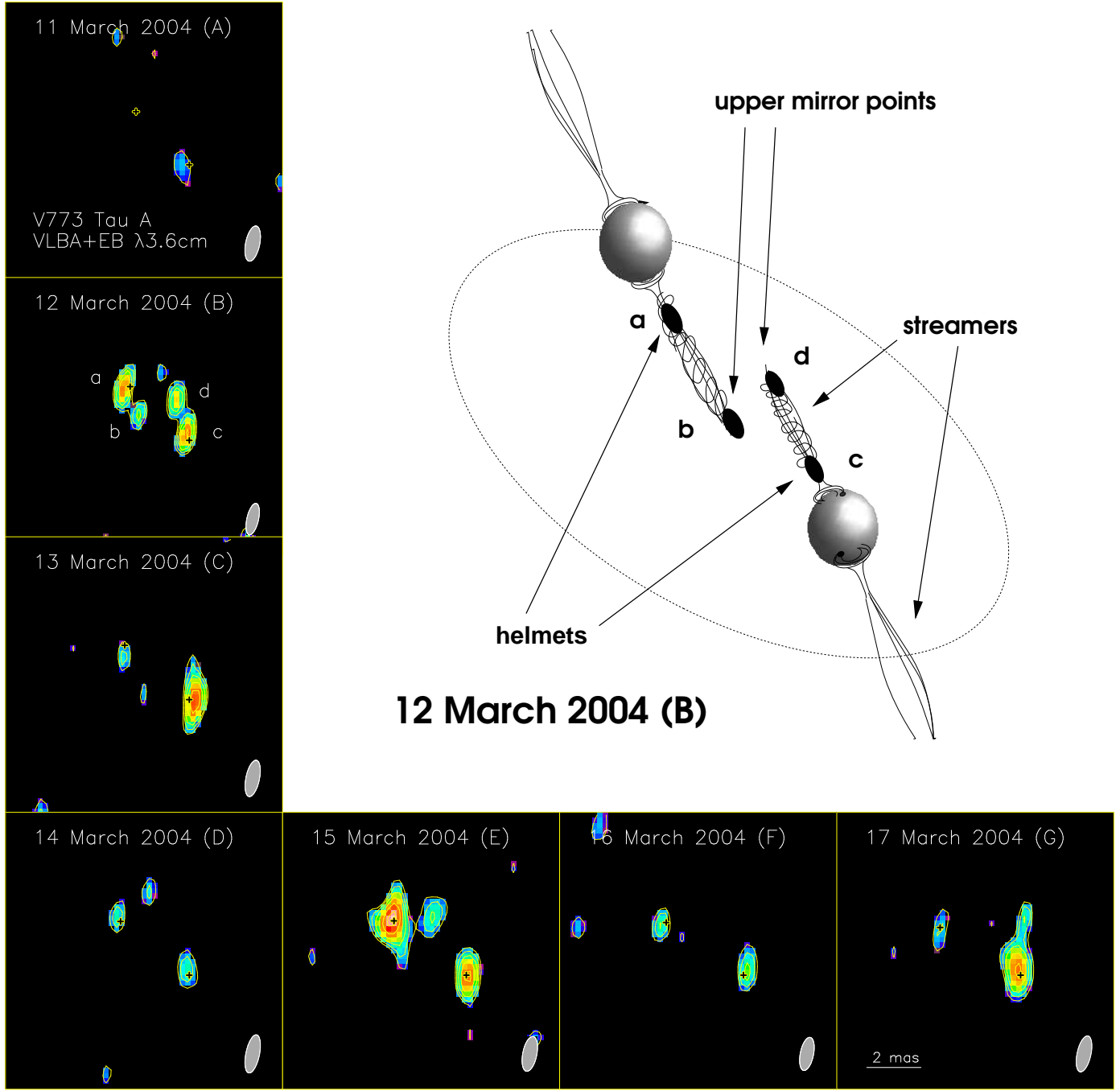}}
\caption{
 Seven consecutive 8.4\,GHz VLBA+EB images of the young stellar binary system V773~Tau A.
Stellar positions are indicated by crosses.
  Starting with panel A, the peak flux density in the images
  are 0.1, 0.5, 0.5, 0.2, 0.7, 0.2, 0.5  mJy/beam.
  Contour levels are $-$1, 1, 1.5, 2, 2.5, 3, 4 and 5 times the
  3 $\sigma$ (0.1 mJy/beam).
  The beam size is shown in the bottom left corner of each panel (average beam
  size: 1.3$\times$0.5 mas).
 The sketch of the system, not to scale, shows 
helmet streamers, i.e. structures formed by a helmet
 located on top  of a coronal loop anchored on a rotating star,
 and a very large streamer. 
The topology is here simplified,
for each star are only drawn two  streamers
of a  possible solar-like streamer belt.
}
\label{vlbi}
\end{figure*}

Besides Fig. ~\ref{vlbi}-A, where only one weak (3$\sigma$) feature
is present, two structures  are always present 
 in all the other images (Fig. ~\ref{vlbi}B-G),
here called North-East, NE, and South-West, SW. The two structures
can be quite complex as in Fig. \ref{vlbi}-B where they show two components
each  (``a", ``b" for the NE and ``c", ``d" for the SW).
The integrated flux measured for the  NE and SW features with the AIPS task JMFIT is
given in Fig.~\ref{eff} along with the total flux measured by
 the Effelsberg 100-m telescope.
As can be seen in Fig.~\ref{eff} the
VLBI flux is very low, i.e. $\leq$1 mJy,
and no significant fraction of circularly and linearly polarized radiation
was detected.
  The higher level of total flux observed with the
  Effelsberg 100-m telescope, with a beam
  of $\sim$ 1.5 arcminutes, 
implies the existence of larger magnetic structures
  around those traced with VLBA+EB at milliarcsecond scales.
The position of  the brightest 
radio peak associated with each NE and 
SW feature 
were determined and their separations are given in the  
4th column of Table 1.

The visual orbit  derived from the
interferometric-spectroscopic orbital model of Boden et al. (2007)
only shows (Fig.~\ref{boden})  the  relative position
between the two
components of the system (i.e. their separation, see 5th column of Table 1), 
it does not automatically identify  the position of
both components on the radio maps.

Comparing the 4th and 5th columns
one can see that the separation between
the two stars grows  from 2.7 mas at run A
to 3.5 mas at run G. 
This is, in fact, consistent with
the growing distance between the two radio peaks as well.
Besides run G, where there is a difference of 
0.4 mas (6 $R_*$)   between the stellar separation and 
the separation of the radio
peaks, all the other distances  agree within their errors
(0.1 mas for both optical and radio results).
Radio peaks and stars are therefore  clearly associated.

Overlapping   the
primary star in
Fig.~\ref{boden}  to the SW feature of 
the radio image with highest signal-to-noise ratio
(i.e. Fig.~\ref{vlbi}-E)
not only
the position of the secondary automatically resulted to be centered 
on the 
brightest radio peak of the NE  feature of that image 
but moreover, the position of the two stars
resulted to be close to the  two radio peaks in all the other images.
In order to avoid a bias on the primary's position
we repeated this procedure overlapping,
this time, the secondary star in Fig.~\ref{boden}
with the NE  feature of  Fig.~\ref{vlbi}-E.
The difference between  
the two determinations of the stellar positions 
in all images was at the most 0.08 mas (1.2 $R_*$). 
The stellar  positions, 
given as a  cross in the radio images of Fig. ~\ref{vlbi},  
were chosen at  the
center between the two determinations.
\begin{table}
\label{table:log}
\begin{center}
\begin{tabular}{ccccc}
\hline
Run     & Date        & $\Phi$& Ha-Hb&Aa-Ab\\
        & (JD)        &       & (mas)& (mas)       \\
    \hline
{\tt A} & 2453076.34 & 0.32 & &2.7    \\
{\tt B} & 2453077.34 & 0.34 &2.8&2.9    \\
{\tt C} & 2453078.34 & 0.35 &3.2&3.1   \\
{\tt D} & 2453079.34 & 0.37 &3.2&3.2   \\
{\tt E} & 2453080.34 & 0.39 &3.4&3.3  \\
{\tt F} & 2453081.34 & 0.41 &3.5&3.4  \\
{\tt G} & 2453082.34 & 0.43 &3.1&3.5  \\
    \hline
    \end{tabular}
    \vspace*{-8pt}
    \end{center}
    %\scriptsize{
    \caption{
    Stellar and helmet separations in the young stellar binary system V773~Tau~A during each VLBA+EB observation. Julian day 2453076.34 corresponds to  March 11, 2004 at 20:10. 
Ha and Hb are the two radio peaks in the VLBI images 
associated with the helmets.
Aa and Ab are the primary  (V773 Tau Aa) and secondary (V773 Tau Ab)
stars  of the system respectively.
The error associated to the  separations
is  $\pm 0.1$mas. 
    At  a distance  of 136 pc,
    one stellar radius, $R_*$,  corresponds to 0.069 mas.
    }
    \end{table}

\subsection{Helmet Streamers}\label{NSXRB}

Whereas on the first day, (March 11, 2004, Fig.~\ref{vlbi}-A), no clear signal is detected,
on the second day, (March 12, 2004, Fig.~\ref{vlbi}-B), two structures
(NE and SW) with a similar morphology are present.
The NE structure  is formed by  two components, ``a" and ``b", separated by 17 R$_*$ 
and the SW one  is formed by  two components, ``c" and ``d", separated by
19 R$_*$.
Both structures are therefore  within the range $15 R_*\leq H \leq 26 R_*$ predicted for the size of the streamers so that 
interbinary coronal collisions can take place.

In all the panels of Fig. ~\ref{vlbi} it is clear that the location of the two stars 
is always close to
the strongest peak of the NE and SW features.
This  coincidence indicates that
 the peaks ``a" and ``c" in  Fig.~\ref{vlbi}-B,
are the helmets of the helmet
streamers,
i.e. the  cusp-shaped field line located on top of a stellar sized
coronal loop.
Two facts indicate that Fig.~\ref{vlbi}-B shows two helmet streamers 
just entering in collision (or immediately after):
{\it i})  
both structures are well distinguished and each  is clearly associated with one of the stars,
whereas 24 hours earlier they were absent (Fig.~\ref{vlbi}-A),
and {\it ii}) the structures are nearly parallel. This is the very first direct evidence revealing that even if the flare origin is magnetic reconnection
due to interbinary collision, both stars emit independently in the radio range with structures of their own.

The general scenario could be the following: each of the two stars has
developed solar-like  helmet streamers and
 due to the  rotation of the two stars (see Sect. 3.3) the streamers approach each other
inducing  a compression of the magnetic field lines and therefore, a successive change of the topology of the two magnetic
configurations. Magnetic reconnection is a possible and probable consequence of this
process as observed for some solar flares  triggered by
interactions between new and older emergences of magnetic flux
in the same area (Nishio et al. 1997; Massi 2007).

The relativistic particles,
produced by magnetic reconnection at the top of the streamer,
spiralling around magnetic field lines emit radio synchrotron  
radiation (Massi et al. 2006).
The particles travel  
 along  the streamer   with low magnetic field strength $B_s$,
 down towards the star to the  helmet (``a" and ``c" in Fig.~\ref{vlbi}-B),
with  magnetic field $B_H$, where $B_{H}>B_s$.
 The helmet will therefore act as a mirror point for the particles with pitch angle
 $\theta > {\rm arcsin} \left ({B_s / B_{H}}\right)^{1/2}$ 
% $\theta > {\rm arcsin} \left ({B_s \over B_{H}}\right)^{1/2}$ 
reflecting them back into the streamer (Melrose $\&$ Brown 1976).
 In the absence of any confinement, i.e. in an open streamer, the particles would leave the structure
 of 17--19 $R_*$ in about
 18 $R_*/{\rm c} \simeq 85$ sec.
 Since
 the emission is observed for a much longer time interval,
at least partial confinement must exist.
Postulating the presence of a second mirror,  far out from the stellar surface
at the top of the streamer,
Massi and collaborators (2006)
modelled  the observed  fading of the emission after a
giant flare
of 360 $\pm$ 17 mJy
as a continuous slow leakage of  emitting
electrons during each reflection between the two mirror points, one close
($\leq 5 R_*$) to the
stellar surface, i.e. the helmet, and the second far out at the top
of the streamer at a distance  of up to $20 R_*$.
The two peaks ``b" and ``d" in Fig.~\ref{vlbi}-B are the observational evidence of
these postulated
 upper mirror points. They are in fact very strong 
due to   the recent  interaction between  the two stars
magnetic  field and  their resulting compression (see Sect 3.4).

\subsection{Colliding Helmet Streamers Anchored on Rotating Stars}\label{geo}

 The image of March 13 (Fig.~\ref{vlbi}-C) cannot be related in a clear way  to the image of March 12 (Fig.~\ref{vlbi}-B).
 The flux density of the SW structure and  the total flux
 observed with the Effelsberg 100-m radio telescope (see  Fig.~\ref{eff}) are higher in Fig.~\ref{vlbi}-C than in Fig.~\ref{vlbi}-B. This means that at the time of the earlier image either a flare had just started,
 and its peak was reached between March 12 and 13
 or,
 another flare occurred within the 24 hours separating the images, and it is seen in its decaying phase in the image of March 13.
On March 14 (Fig.~\ref{vlbi}-D) there are traces of a new  upper mirror point
in the NE
indicating  that a new flare has occurred before this observation.
There are therefore 2-3  flares in 3 days, i.e. a flaring rate of up
to one flare per day.
How does this fit with the stellar rotation?

Rydgren \& Vrba (1983) 
monitored  the system with {\it UBVRI} photometry and observed 
light  variations with a period of about 3 days.
Besides, 
using the radii of the two stars, $R_{\rm primary}$=2.22 $R_{\odot}$ and  $R_{\rm secondary}$=1.74 $R_{\odot}$ (Boden et al. 2007),
  their projected rotational velocities, 41.4 km s$^{-1}$ and 41.9 km s$^{-1}$ (Welty  1995), assuming that the spin axes of the stars are parallel to
 the axis of the orbit and using the orbital inclination angle  $i=65.9^{\circ}$, we obtain a  rotational period of 2.48 days and 1.92 days for the primary and secondary respectively.

%Taking into account the deduced flaring rate and the stellar rotational period we suggest a possible configuration where streamers are located
%around the star (see sketch in Fig.~\ref{vlbi}).
A geometry where the  spin axes of the stars are  parallel to
the axis of the orbit, implies interbinary collisions only if
the streamers are located around the equator. This is
likely to be the case, since 
 in the Sun helmet streamers are located on the equator during the solar
minimum of activity and  in  all the latitudes during the
 maximum of activity (Gibson 2001).

Therefore, following the solar analogy, it is clear that
the  helmet streamer configuration around each
star in \vt  could be  more complex  
than what is  shown in  the  sketch of  Fig.~\ref{vlbi}.
This can also be inferred from the fact that
a rotational period of  1.92 days for the secondary
implies a rotation of about 180$\degr$/day.
Assuming the simplest configuration of Fig.~\ref{vlbi}, this rotation would imply
that an emitting helmet-upper mirror point structure should always appear in the
same position angle since all the images were taken every 24 hours.
Instead  the position angles in Figs.~\ref{vlbi}-B, D, and  E,
are   208$\degr$, 308$\degr$ and, 273$\degr$, respectively (North throw East).
A  solar-like  streamer belt giving more than 1-2 flares per rotation 
could be a straightforward explanation.  
Only a new set of observations lasting
longer than  5.5. hours each will lead to a better understanding of this complex
topology.
\subsection{Short-Living Upper Mirror Points}\label{geo}

It is worth noting that the flux density in the  VLBA+EB images never reaches even 50$\%$ of
the total flux density observed with the Effelsberg 100-m telescope with a beam
of $\sim$ 1.5 arcminutes (see Fig. 3).
Most of the flux is therefore emitted from  electrons that are trapped in
magnetic structures which are clearly larger than the milliarcsecond features
imaged by the VLBA+EB.
On March 11 we observed just one helmet and a day later two complete helmet streamers were present. Nevertheless, on both days the Effelsberg
100-m telescope measured
the same total flux density of 2.4 mJy. However, whereas on March 11 only 14$\%$ of this flux is in the VLBA+EB
image, on the image of March 12 there is a 46$\%$ of the total flux.
This seems to indicate that the  upper mirror points
gradually relax on larger structures
 which are still confining the emitting electrons but are not ``visible" anymore at the milliarcsecond scales of the VLBI.

\begin{figure*}[htb]
\centering
\resizebox{\hsize}{!}{\includegraphics[angle=-90, scale=0.3]{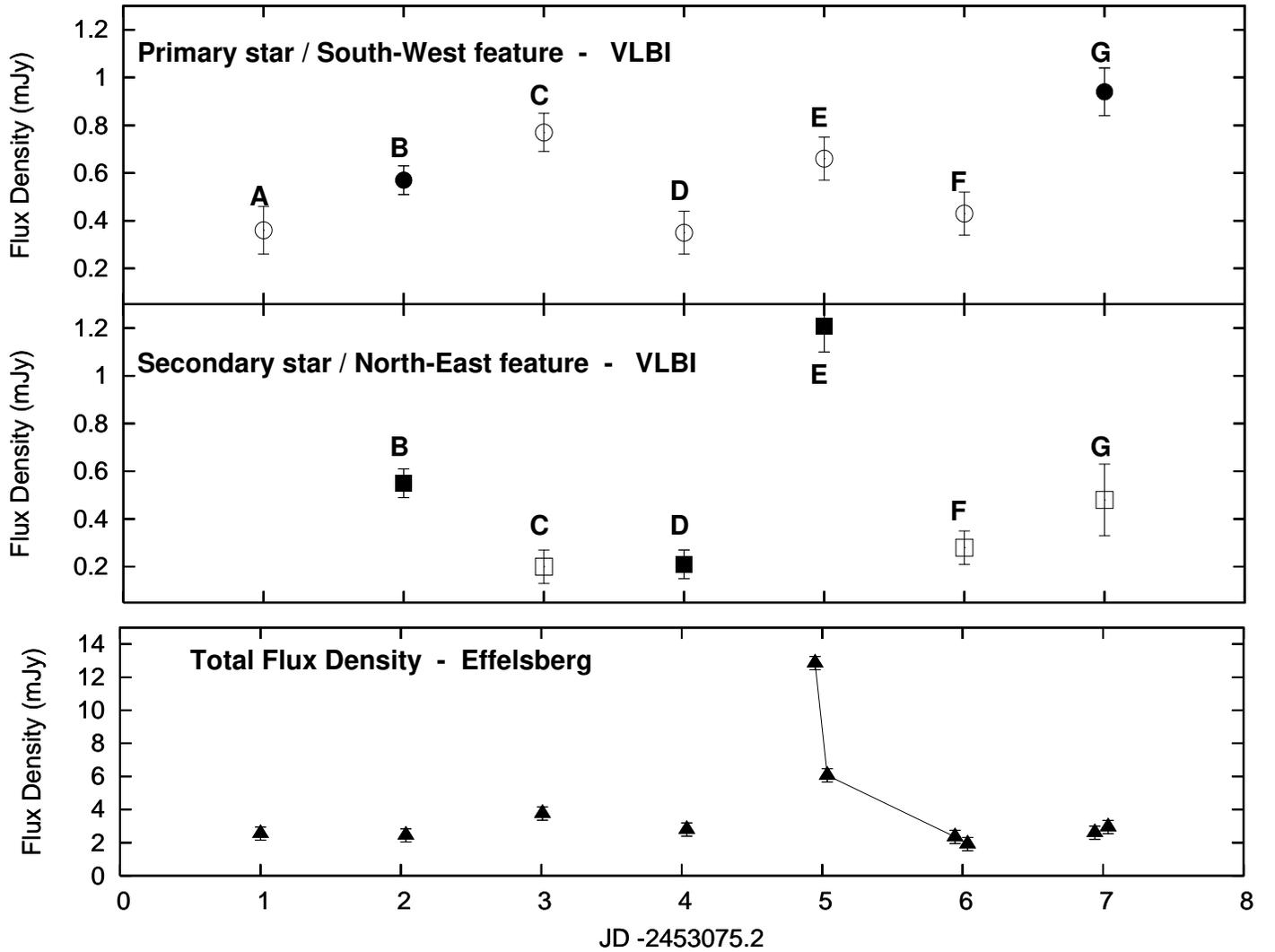}}
\caption{
Flux density versus time.
 Top panel: flux density of the SW radio structures (associated with the primary star)
  in the  VLBA+EB maps. Filled circles express the combined flux
  of  helmets and upper mirror points; empty circles indicate that only helmets are present.
  Middle panel: flux density of the NE radio structures (associated with the secondary  star).
  Filled squares when helmets and upper mirrors are both present, empty squares when only helmets are present.
  Bottom panel: total flux density measured with the Effelsberg 100-m telescope outside the
  VLBA+EB observing time.
  The points which are likely to belong to the same flaring event have been connected
  (days 5-6).
}
\label{eff}
\end{figure*}

On March 15 (Fig.~\ref{vlbi}-E) one helmet and its related  upper mirror point
   can be clearly distinguished in the NE structure but
   only the helmet is present in the SW one.
Thanks to the monitoring at the Effelsberg
100-m telescope we know that at the time of this image
   the flare is already in its decaying phase.
   In fact,
   while a total flux density of 12 mJy was measured with the Effelsberg 100-m telescope
   two hours before
   the  VLBA+EB observation,
   only 6 mJy were measured at the beginning of this observation (Fig.~\ref{eff}).
It is significant that  in Fig.~\ref{vlbi}-B, that is just after/during
a collision, the sources are point-like, whereas
in   Fig.~\ref{vlbi}-E, 
showing an already evolved situation, the sources are extended.
Relativistic
electrons which diffuse from the relaxing upper mirror points remain
trapped in progressively larger structures, intermediate between the
 VLBA+EB milliarcsecond
scale and the Effelsberg 100-m telescope arcminute scale.
It is worth to note that in the flare decay scenario of Fig.~\ref{vlbi}-E
the NE elongated structure on the secondary star is not parallel to the  helmet streamers seen in the B run but clearly rotated.
%A  rotation $\leq 90 \degr$ implies an elapsed time
% $\leq 1.92/4$ days or $\leq$ 11 hour.
There seems to be three facts indicating that time has elapsed from the
streamer-streamer collision: the clear rotation of the streamers with respect to the
line joining the two stars,
their non point-like appearence, and  the eventual absence of upper mirror point(s).
With this respect, in the last observation of March 17
(Fig. 2-G), the facts that on one hand the helmet streamer anchored on the 
primary
star appears substantially rotated and extended
and  on the other hand, there is a lack of the upper mirror in the NE
structure, suggest a flare decay scenario as described for Fig. 2-E.

The fact that after a certain time interval  from the occurrence
of a flare  only the helmet tend to remain
implies that its corresponding  upper mirror point
either already relaxed 
to a larger structure  or that an upper mirror
 point might not even have been  created.
 In both cases, the result is that the upper mirror point is clearly
not a permanent feature.

What could be the  origin of these short-living
upper mirror points at distances of tens of stellar radii?
Up to now there is no known solar analogy to these upper mirror points in the solar streamers.
On other hand, it is known that 
the solar rotation pulls the magnetic field lines into a spiral shape
(Parker 1958, 1969)
and a much stronger effect could be expected 
in \vt due to the faster rotation of its stellar components
(more than one order of magnitude higher than the solar rotation).
The faster rotation could lead to strongly curved  streamers anchored on  
the same star to interact with  each other.
The feasibility of such an interaction can  be  verified 
by computing the  Alfv\'en radius,   that is the stellar distance 
where the magnetic field lines become to be 
curved  by  star rotation (Andr\'e 1988):
\beq
{R_A \over R_*}= 26 \left ({B_*\over 10^4 \rm{G}} \right) ^{{1 \over 3}}
\left ( {\dot{M} \over 10^{-10} \rm{M}_\odot ~
\rm{yr}^{-1} } \right )^{-{ 1 \over 6}} \left ( {{\rm v}_{\infty} \over 10^3 ~ \rm{km ~ s}
^{-1}} \right )^{{1 \over 6}} \left ({P \over 1 \rm{d}} \right)^{{ 1 \over 3}}.
\eneq
The result is that
for a mass loss of  $10^{-13} \rm{M}_{\odot}~  \rm{yr}^{-1}$ 
(Guenther \& Emerson 1997),
a terminal velocity of a  few hundred kilometers per second,
 $P\simeq2$ d,  and
 a  magnetic field strength at the stellar surface of 2600 G
(Bower et al. 2003),
 the  Alfv\'en radius is  $\sim 60 R_*$.
This value  that clearly exceed the 
observed upper mirror points position
rules out this self-interaction  as a possible explanation
and leaves as the only alternative the
fact that interbinary collision 
produces the upper mirror points.

\section {Conclusions}

\subsection{Halo and Giant Flares}\label{geo}
The complete extent of the helmet streamers
is  the observed value of   18 $R_*$, between helmet and upper mirror point, plus the distance
      of the helmet itself from stellar centroid.
The  flare activity 
observed on March 17 when the orbital separation was 48 $R_*$
implies a complete extent of $\sim 24 R_*$
and therefore   helmets located on top of coronal loops
of  $\sim$ 5 $R_*$ as expected on the basis of 
the theoretical predictions
(see Sects. 1 and 3.2).
      This explains the practically permanent flaring activity in V773 Tau A,
from a level of a few mJy  around apoastron (52 $R_*$) 
to more than  100 mJy  at periastron (30 $R_*$) (Massi et al. 2002).
At the short periastron  distance  the two coronae of  $\sim 24 R_*$   
 overlap, giving rise to  the 
observed giant flares (i.e. up to 360 mJy at $\lambda$ 3mm,  Massi et al. 2004).
      The result of such a situation,
      where the whole streamer participates
       in the magnetic reconnection process, is
 a large structure, which holds the emission peaks at the position of the
       two helmets anchored on each of the two stars
      as in the
VLBI image of Phillips and collaborators (1996) discussed in Sect. 1.

\subsection{Summary}\label{geo}
To what extent can our knowledge of the Sun guide us to understand other stars?
Here we have reported the discovery of
solar-like helmet streamers  anchored on the two stars of a very young binary system, V773 Tau A.
 The extended helmet streamers, observed at radio wavelengths with very high resolution, cover
 the interbinary distance. During the stellar rotation these helmet streamers come into  collision  producing  periodical
 flares.
Electrons,  relativistically accelerated in the collision,
remain trapped in the  helmet streamers and emit
synchrotron radiation observed in the radio band.

  The enigmatic
magnetic semi-open structures typical of helmet streamers, until now observed only on  the
Sun, have been
postulated to play
a key role not only in the  processes of coronal mass ejection and
slow solar wind (Wiegelmann et al. 2000; Schwenn 2006),
formation of prominences at large height 
in rapidly rotating stars (Jardine \& van Ballegooijen 2005),
but also in the processes 
of jet formation in  microquasars/AGN (de Gouveia dal Pino 2005)
 and outflows in young stellar objects (Ostriker \&  Shu 1995; Ferreira et al.
2006).
%%Their discovery in other stars than the Sun, at a wavelength easy to
%%be observed at high resolution, will therefore lead to
Their discovery in stars other than the Sun, at a wavelength
observable at high resolution, will therefore lead to
a deeper understanding of these important physical processes.

\begin{acknowledgements}
%\begin{acknowledegments}
Based on observations with the  Effelsberg 100-m
telescope of the
Max-Planck-Institut f\"ur Radioastronomie (MPIfR) and
the Very Long Baseline Array (VLBA).
The   VLBA is a   facility of the National Radio Astronomy Observatory (NRAO),
operated by
Associated Universities Inc. under a cooperative agreement with the National Science Foundation
(NSF).
We thank Peter L. Biermann, Nicolas Grosso, Giannina Poletto,
and  Thomas Preibisch 
for interesting  suggestions.
We are grateful to the referee Manuel G\"udel
 for suggestions and comments which helped clarifying 
our manuscript on several important points.
\end{acknowledgements}

%\begin{figure*}[]
%\centering
%\resizebox{\hsize}{!}{\includegraphics[angle=0, scale=0.75]{fig2rot.eps}}
%\caption{  WITH ROTATION.  Could this be  an alternative to Fig.2 ?. Please comments/suggestions?
%}
%\end{figure*}[]


\begin{thebibliography}{}

\bibitem{black} Andr\'e, P., Montmerle, T., Feigelson, E. D., Stine, P. C., \& Klein, K.  1988, \apj,  335, 940
%\bibitem{black} Aschwanden, M.  2006,  \ssr,  124, 361
\bibitem[Bo]{} Boden, A.~F., Torres, G., Sargent, A.~I., et al. 2007, \apj,
 670, number 2 in press, astro-ph/0706.2376
%\bibitem[Bo]{} Boden, A.~F., Torres, G., Sargent, A.~I., et al. 2007, \apj, in press, astro-ph/0706.2376
\bibitem{black} Bower, G.~C., Plambeck, R.~L., Bolatto, A., et al. 2003,  \apj,  598, 1140
\bibitem[5)]{} de Gouveia dal Pino, E.~M.\ 2005, Magnetic Fields in the Universe: From Laboratory and Stars to Primordial Structures., AIPC, 784, 183
\bibitem{black} Feigelson, E.~D., Welty, A.~D., Imhoff, C., et al. 1994,  \apj,  432, 373
\bibitem{black} Ferreira, J., Dougados, C., \& Cabrit, S.  2006, \aap, 453, 785
\bibitem{black} Gibson, S.~E.  2001,  \ssr,  97, 69
\bibitem{black} Guenther, E.~W., \& Emerson, J.~P. 1997,  \aap, 321, 803 
\bibitem{a} Jardine, 
M., \& van Ballegooijen, \aap 2005, \mnras, 361, 1173 
\bibitem{black}
Massi, M.\ 2007, Memorie della 
Societa Astronomica Italiana, 78, 247
\bibitem{black}
Massi, M., Menten, K.,  \&  Neidh\"ofer J.  2002, \aap,  382, 152
\bibitem{black}
Massi, M., Forbrich, J., Menten, K. M., et al.  2006, \aap,  453, 959
\bibitem{black} Melrose, D. B. \& Brown, J. C. 1976, MNRAS, 176, 15
\bibitem{7}
Nishio, M., Yaji, K., Kosugi, T., Nakajima, H.  \& Sakurai, T.  1997, \apj, 489, 976
\bibitem{black} Ostriker, Eve C., \&  Shu, Frank H.  1995, 	\apj,  447, 813
\bibitem{ApJ} Parker, E.~N.\ 1958, \apj, 128, 664 
\bibitem{S} Parker, E.~N.\ 1969, Space Science Reviews, 9, 325 
\bibitem{black} Phillips, R. B., Lonsdale, C. J., Feigelson, E. D. \&  Deeney, B. D.  1996, \apj,  111, 918
\bibitem{rv}
Rydgren, A. E. \& F. J.  Vrba, F. J.
1983,  \apj,  267, 191
\bibitem{black} Schwenn, R.  2006, Space Science Rev.,  124, 51
\bibitem{black}
Skinner, S. L., Guedel, M., Koyama, K. \&  Yamauchi, S.
1997,  \apj, 486,  886
\bibitem{black} Suess, S. \&  Nerney, S.  2004,  Advances in Space Research, 33,  668
\bibitem{black} Tsuboi, Y., Koyama, K., Murakami, H., Hayashi, M., Skinner, S., \&  Ueno, S.  1998, \apj,  503, 894
 \bibitem{black}
Vourlidas, A., 2006,  Solar Activity and its Magnetic Origin.  eds Bothmer, V. \&  Hady, A.  Cambridge University Press, p.197
\bibitem{black}
Welty, A.D. 1995, \aj, 110, 776
\bibitem{black}
Wiegelmann, T., Schindler, K., \& Neukirch, T.  2000,  Sol. Phys.,  191, 391
\end{thebibliography}
\end{document}